\documentclass[aps,pra,twocolumn,tightenlines,groupedaddress,showpacs]{revtex4-1}
\usepackage{epsfig}
\usepackage{amsmath}
\usepackage{amsfonts}
\usepackage{color}
\usepackage{graphicx}
\usepackage{subfigure}

\def\bra#1{{\langle#1|}}
\def\ket#1{{|#1\rangle}}

\def\expect#1{{\langle#1\rangle}}

\def\H{{\hat H}}

\def\L{{\hat L}}
\def\Ldag{{\hat L}^\dagger}

\def\Op{{\hat O}}
\def\id{{\hat I}}
\def\Z{{\hat Z}}
\def\p{{\hat p}}
\def\Sx{{\hat S}_x}

\def\Sz{{\hat S}_z}

\def\R{{\hat R}}
\def\sigmaX{\hat{\sigma}_X}
\def\sigmaZ{\hat{\sigma}_Z}
\def\DZ{\expect{\Delta\Z^2}}
\def\Dp{\expect{\Delta\p^2}}
\def\DR{\expect{\Delta\R}}
\def\DZcube{\expect{\Delta\Z^3}}

\def\DQ{\expect{\{\Delta\Z , \Delta\p^2 \} } }
\def\DS{\expect{\{ \Delta\Z^2, \Delta\p \} } }

\begin{document}
\title{Gaussian approximation and single-spin measurement in OSCAR MRFM with spin noise}

\author{Shesha Raghunathan}
\email{sraghuna@usc.edu}
\author{Todd A. Brun}
\email{tbrun@usc.edu}
\affiliation{Center for Quantum Information Science and Technology,
Communication Sciences Institute, Department of Electrical Engineering, 
University of Southern California, Los Angeles, CA 90089, USA.}
\author{Hsi-Sheng Goan}
\email{goan@phys.ntu.edu.tw}
\affiliation{Department of Physics and Center for Theoretical Sciences,
National Taiwan University, Taipei 10617, Taiwan}
\affiliation{Center for Quantum Science and Engineering,
National Taiwan University, Taipei 10617, Taiwan}

\begin{abstract}
A promising technique for measuring single electron spins is magnetic resonance force microscopy (MRFM), in which a microcantilever with a permanent magnetic tip is resonantly driven by a single oscillating spin.  If the quality factor of the cantilever is high enough, this signal will be amplified over time to the point that it can be detected by optical or other techniques.  An important requirement, however, is that this measurement process occur on a time scale short compared to any noise which disturbs the orientation of the measured spin.  We describe a model of spin noise for the MRFM system, and show how this noise is transformed to become time-dependent in going to the usual rotating frame.  We simplify the description of the cantilever-spin system by approximating the cantilever wavefunction as a Gaussian wavepacket, and show that the resulting approximation closely matches the full quantum behavior.  We then examine the problem of detecting the signal for a cantilever with thermal noise and spin with spin noise, deriving a condition for this to be a useful measurement.
\end{abstract}


\maketitle

%\\\\\\\\\\\\\\\\\\\\\\\\\\\\\\\\\\\\\\\\\\\\\\\\\\\\\\\\\\\\\\\\\\\\\\\\\\\\\\\\\\\\\\\\\\\\\\\\\\\\\\\\\\\\\\\\\\\\\\\\\\\\\\\\\\\\\\\\\\\\\\\\\\\\\\\\\\\\\\\\\\\\\\\\\\\\\\\\\\\\\\\\
%
\section{Introduction}
%
% 1. Single qubit measurement is fundamental to QIP with spin based systems  
% 2. Other applications include: molecular imaging, protein mapping etc.
% 3. Since the proposal was first made in 1991, there has been considerable development in the experimental domain which culminated in the detection of single-qubit in 2004. Formal understanding of its operation from quantum level upwards was missing.
% 4. Contributions of our work are two fold: (i) modeling and simulation of the MRFM system in (stochastic) master equation setting, and, (ii) gaussian approximation of the cantilever. 
%
Sidles, in 1991, first proposed the use of magnetic resonance and mechanical oscillators to sense a weak force~\cite{Sidles91}. Ever since this proposal, there has been considerable progress in the experimental implementation of such a technique~\cite{mrfm-experimental, Stipe01, Rugar04}, culminating in single spin detection in 2004~\cite{Rugar04}. That the force detected is of the order of attonewtons, highlights the usefulness of magnetic resonance force microscopy (MRFM) based techniques.

MRFM based techniques have a wide range of applications. One such application is imaging; in fact, imaging was the original motivation for proposing MRFM; imaging at the nanoscale like that of biological molecules such as  proteins, viruses etc., is of immense value to the society at large~\cite{Degen09}. Another application, and the one that is of interest to us, is the use of MRFM for single spin measurements. Numerous models of quantum computing involving spins have been proposed~\cite{spin-models}, and these models require the ability to do single spin measurements, in addition to purely scientific value of such a measurement capability.  

% Discuss theoretical developments in explaining MRFM based single spin measurement
%
In 2001, Stipe et al.~\cite{Stipe01} first implemented the OScillating Cantilever-driven Adiabatic Reversals (OSCAR) protocol. The frequency shift that takes place in this protocol, can be measured experimentally with high precision. The idea of this spin manipulation protocol is to transform the cantilever-spin interaction force into a shift in the resonant frequency of the oscillating cantilever, by using a gain-controlled feedback mechanism; the interaction force between the cantilever and the spin, which is either attractive or repulsive depending on the orientation of the spin, gets transformed to a positive or a negative shift in frequency; by measuring this shift one can determine the orientation of the spin.

Theoretical analysis quickly followed with a series of work by Berman et al.~\cite{Berman02, Berman03, Berman04, Berman06} and Brun and Goan~\cite{Brun03, BrunGoan05}. Berman et al.~\cite{Berman02, Berman03, Berman04, Berman06} use a semi-classical approach in their analysis, wherein they treat cantilever as a classical Harmonic oscillator while treating the spin quantum mechanically. Brun  and Goan~\cite{Brun03, BrunGoan05} provide a fully quantum description of the system, in that they treat both cantilever and spin quantum mechanically; they also consider decoherence effects acting on the cantilever in their analysis, including continuous measurement of the cantilever due to optical interferometric techniques. Though modeling and analysis of the system using a fully quantum description is an important theoretical development, Brun and Goan~\cite{Brun03, BrunGoan05} describe the system as a pure quantum state, instead of a (more general) density matrix, and use quantum (pure) state diffusion techniques to describe their evolution, instead of a stochastic master equation evolution~\cite{Trajectories}; also, they do not consider decoherence effects acting on the spin. 

In this work, we analyze the MRFM system in generality within a Markovian framework. We describe the system using a density matrix, and evolve it using a stochastic master equation~\cite{Trajectories}. We consider decoherence effects acting on both the cantilever and the spin. The stochastic master equation, though it gives a fully quantum-mechanical description of the system, is numerically expensive. To ameliorate this, we show that the state of the cantilever can be approximated as a Gaussian wave packet, leading to a description of the MRFM system by a closed set of $11$ coupled stochastic differential equations. We show that the Gaussian approximation is valid in the parameter regime of interest, by numerically comparing the evolution of the Gaussian equations with the fully quantum stochastic master equation. 

Further, we use the Gaussian equations to analyze the OSCAR protocol as a tool for single-spin measurement. We consider the constraints set by the spin noise on a single-spin measurement. For the parameter values chosen, we calculate a bound on the spin noise time scale. We choose a rate for spin noise that satisfies this time scale, and numerically show that the OSCAR protocol can indeed be used to do single-spin measurement.

\subsection{Overview of the paper}
Sec.~\ref{sec:model_and_operation} gives an overview of the MRFM system and the OSCAR protocol. In Sec.~\ref{sec:eff_Hamiltonian}, we use the adiabatic approximation and derive an effective Hamiltonian for the OSCAR MRFM system. We consider the decoherence effects acting on both the cantilever and the spin, in Sec.~\ref{sec:modeling}; the sources of decoherence include the thermal bath, continuous measurement of the cantilever using optical interferometry, and spin noise due to magnetic sources. We then discuss the feedback mechanism used to implement the OSCAR protocol (Sec.~\ref{sec:feedback}). We derive equations for the moments in Sec.~\ref{sec:moment_eqns}; we consider the unitary case first (Sec.~\ref{sec:unitary}), before proceeding to the more general open system evolution (Sec.~\ref{sec:open_system}). Finally, we present the numerical results in Sec.~\ref{sec:results}. We first discuss the parameter values that are used in our simulations; we consider the effectiveness of the Gaussian approximation by comparing it to the fully quantum stochastic master equation evolution (Sec.~\ref{sec:results-gaus}); and lastly, we ask what it takes for an OSCAR MRFM system to be an useful single-spin measurement device.

%\\\\\\\\\\\\\\\\\\\\\\\\\\\\\\\\\\\\\\\\\\\\\\\\\\\\\\\\\\\\\\\\\\\\\\\\\\\\\\\\\\\\\\\\\\\\\\\\\\\\\\\\\\\\\\\\\\\\\\\\\\\\\\\\\\\\\\\\\\\\\\\\\\\\\\\\\\\\\\\\\\\\\\\\\\\\\\\\\\\\\\\\
%
\section{OSCAR MRFM - Model and Operation}
\label{sec:model_and_operation}
\begin{figure*}[htp]
  \begin{center}
    	 \includegraphics[width=4.25in,height=2.75in]{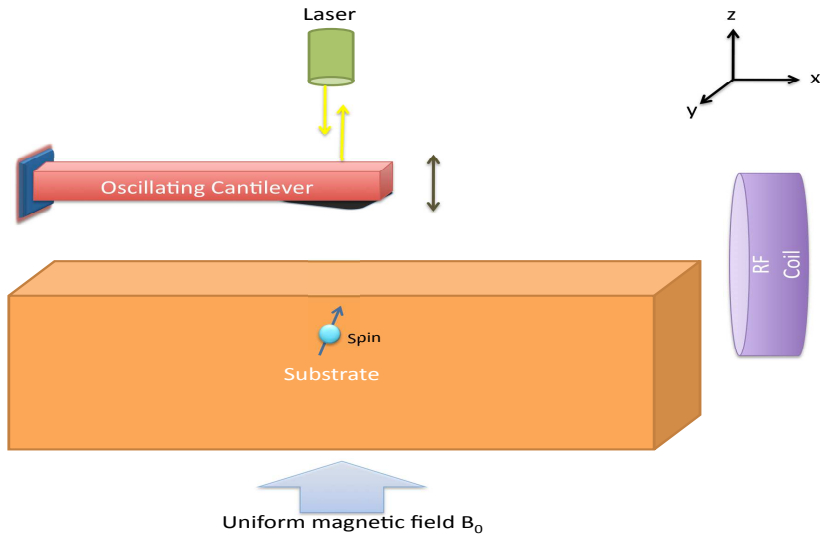}
  \end{center}
  \caption{(Color online.) MRFM system: model and operation. A cantilever with a ferromagnetic tip oscillates in close proximity to a substrate with a free spin. A uniform magnetic field $B_0$ is applied in the $z$ direction and a microwave (RF) field is applied in the $x$-$y$ plane. The position of the cantilever is monitored continuously by shining a laser on the cantilever tip and using optical interference to make measurements. A feedback mechanism (not shown in the figure) maintains the amplitude of the cantilever at a pre-determined fixed value, making the cantilever a frequency-determining element. The direction of the frequency shift reveals the orientation of the spin. }
  \label{fig:mrfm_model}
\end{figure*}

The basic model of MRFM involves a cantilever oscillating in close proximity to a spin that is contained in a substrate (Fig.~\ref{fig:mrfm_model}). A ferromagnet that is attached at the tip of the cantilever interacts with the spin. This interaction changes the amplitude and frequency of the oscillator, which is measured to determine the orientation of the spin.

There are three different sources of magnetic field present in the
system: $(i)$ a constant field, $B_0$, in the $z$ direction, $(ii)$ an
oscillating {\it ac} microwave (RF) field, $B_{ac}$, that is applied in the $x-y$ plane, and, $(iii)$ a dipole field due to the ferromagnetic tip, $B_d$.

The constant field $B_0$ polarizes the spin in the $z$ direction. The
rotating microwave (RF) field, together with the dipole field from the ferromagnet, causes the spin precession axis to periodically reverse with the same frequency as the cantilever. When in resonance, the spin oscillates with the same frequency as that of the cantilever. Under the resonant condition, the force due to interaction of the ferromagnet at the tip of the cantilever and that with the magnetic moment of the spin gets amplified: this amplification leads to a change in the resonant amplitude and frequency of the cantilever. This change can be observed by monitoring the cantilever, usually through optical interference.

In the OSCAR protocol, the oscillation of the cantilever (ferromagnetic tip) causes adiabatic reversals of the spin, which in turn interacts with the cantilever to change the resonant amplitude and frequency of the cantilever~\cite{Stipe01}. In this protocol, feedback is used to shift the resonant frequency of the cantilever; the change in resonant frequency---positive or negative---of the cantilever is used to determine the orientation of the spin. The feedback mechanism used is positive gain controlled, and is used to maintain the cantilever amplitude at a pre-determined fixed value. The cantilever thus behaves as a frequency determining element~\cite{Albrecht91}.

%\\\\\\\\\\\\\\\\\\\\\\\\\\\\\\\\\\\\\\\\\\\\\\\\\\\\\\\\\\\\\\\\\\\\\\\\\\\\\\\\\\\\\\\\\\\\\\\\\\\\\\\\\\\\\\\\\\\\\\\\\\\\\\\\\\\\\\\\\\\\\\\\\\\\\\\\\\\\\\\\\\\\\\\\\\\\\\\\\\\\\\\\
%
\section{Effective Hamiltonian using the adiabatic approximation}
\label{sec:eff_Hamiltonian}
The cantilever and spin form the two parts of the system, and are described by their respective Hamiltonians: 
\begin{equation}
\begin{array}{c}
\H_C(t) \, = \, {\p^2}/{2 m} \: + \: {m \omega^2 \Z^2}/{2} \: - \: f(t) \, \Z \\
\H_S(t) \, = \, \epsilon \, \Sx \, - \, \eta \, \Z \Sz \\
\end{array}
\label{eqn:original_hamiltonian}
\end{equation}
where $\H_C(t)$ and $\H_S(t)$ are the Hamiltonians of the cantilever and spin, respectively~\cite{Berman04}; $f(t)$ is the positive gain controlled feedback mechanism that implements the OSCAR protocol (refer Sec.~\ref{sec:feedback}); and, $\epsilon$ and $\eta$ represent the strength of the microwave (RF) field and spin cantilever interaction, respectively. Note that we have included the interaction term $\eta \, \Z \, \Sz$ as part of the spin Hamiltonian $\H_S(t)$.

In the parameter regime of the OSCAR protocol, the time scales of the spin and the cantilever are well separated. The frequency of spin precession is much greater than the oscillation frequency of the cantilever. On the time scale of the spin, the cantilever part in the interaction term in the Hamiltonian,
\[
\H_I \: = \: - \, \eta \Z \Sz
\]
can be treated as a constant; thus, the quantum cantilever position term, $\Z$, on the time scale of the spin can instantaneously, in the adiabatic limit,  be treated as a classical function $Z$. 

The time dependent spin Hamiltonian in the adiabatic limit is given by
\begin{eqnarray}
\H_S(t) \: = \: \epsilon \, \Sx \: - \: \eta \, Z \, \Sz, 
\end{eqnarray}
whose instantaneous eigenstates are
\begin{eqnarray}
\ket{v_{\pm}(t)} \: &=& \: \frac{\epsilon}{\sqrt{\epsilon^2 + (\pm \lambda + \eta Z)^2}} \,  \ket{\uparrow} \:  \nonumber \\
&& \:\: \:\: \:\: \:\: \:\: \:\: + \:\: \frac{\pm \lambda + \eta Z}{\sqrt{\epsilon^2 + (\pm \lambda + \eta Z)^2}} \,  \ket{\downarrow},
\label{eqn:inst_eigenstate}
\end{eqnarray}
with eigenvalues 
\begin{equation}
\lambda(t) = \sqrt{\epsilon^2 \, + \, \eta^2 \, Z^2}. 
\label{eqn:eigenvalues}
\end{equation}
If the initial state of the spin is
\begin{equation}
\ket{\psi_S(0)} \: = \: \alpha(0) \, \ket{v_{+}(0)} \: + \: \beta(0) \, \ket{v_{-}(0)},
\label{eqn:psi_0}
\end{equation}
then the state of the spin at later times can be written as
\begin{equation}
\ket{\psi_S(t)} \: = \: \alpha(t) \, \ket{v_{+}(t)} \: + \: \beta(t) \, \ket{v_{-}(t)},
\label{eqn:psi_t}
\end{equation}
where $\ket{v_{\pm}(t)}$ is the instantaneous eigenstate of the time-dependent spin Hamiltonian at time $t$. The time evolution of the state is given by the Schr\"{o}dinger equation
\begin{equation}
\frac{d}{dt} \ket{\psi_S(t)} \: = \: -\left( \frac{i}{\hbar} \right) \, \H_S(t) \, \ket{\psi_S(t)}.
\label{eqn:schrodinger}
\end{equation}
Using Eq.~(\ref{eqn:psi_t}) in Eq.~(\ref{eqn:schrodinger}), and solving for $\alpha(t)$ and $\beta(t)$ using the adiabatic approximation, we get 
\begin{eqnarray}
\alpha(t) \: &=& \: \alpha(0) \, \exp\left\{-\frac{i}{\hbar} \int_0^t \, \lambda(s) \, ds \right\}, \, 
		\label{eqn:prob_alpha_t} \\ 
\beta(t) \: &=& \: \beta(0) \,  \exp\left\{\frac{i}{\hbar} \int_0^t \, \lambda(s) \, ds\right\}. 
\label{eqn:prob_beta_t}
\end{eqnarray}
The probabilities of the spin being up or down in the reference frame
of the effective magnetic field are constants of motion, and are given
by $| \alpha(0) |^2$ and $| \beta(0) |^2$, respectively. In the
solution above, we have used the fact that the time scale of
cantilever motion is much slower than the time scale of spin
precession, i.e., the rate of change of the cantilever 
$|dZ/dt| \ll (\epsilon^2/\eta)$, and consequently, we have ignored terms proportional to $dZ/dt$ in deriving Eqs.~(\ref{eqn:prob_alpha_t}) and (\ref{eqn:prob_beta_t}).

Assuming the cantilever and spin are initially decoupled, the initial state of the cantilever-spin system is
\begin{equation}
\ket{\psi(0)} \: = \: \ket{\psi_C(0)} \otimes \ket{\psi_S(0)}, 
\end{equation}
where $\ket{\psi_C(0)}$ is the initial state of the cantilever, and $\ket{\psi_S(0)}$ is the initial state of the spin as defined in Eq.~(\ref{eqn:psi_0}). Further, we assume that the cantilever is localized, and consequently the motion of the cantilever can be described by two wave packets---one corresponding to each spin orientation.

As we know, $\ket{v_{\pm}(t)}$ are the eigenstates of the spin Hamiltonian, and satisfy
\[
\H_S(t) \, \ket{v_{\pm}(t)} \: = \: \pm \lambda(t) \,  \ket{v_{\pm}(t)},
\] 
where $\lambda(t)$ is its eigenvalue [Eq.~(\ref{eqn:eigenvalues})]. We assume that the microwave (RF) field, $\epsilon$, is much stronger than the interaction strength between the cantilever and the spin $\eta Z$. Expanding $\lambda(t)$ as a binomial series, we get
\begin{equation}
\lambda(t) \: = \: \epsilon \left[ 1 + (\eta Z/\epsilon)^2/2 + O((\eta Z/\epsilon)^3) \right].
\label{eqn:lambda}
\end{equation}
As the microwave (RF) field dominates the interaction strength, we can ignore third---and higher order---contributions in this expansion. Since the first term in the expansion is a constant, it amounts to a phase factor, and can be ignored as well.

Let us now consider the action of the total Hamiltonian on the joint cantilever and spin states:
\begin{equation}
(\H_C + \H_S(t)) \, (\ket{z} \otimes \ket{v_{\pm}(t)}) \: = \: (\H_C \pm \lambda(t)) \, (\ket{z} \otimes \ket{v_{\pm}(t)}),
\end{equation}
where $\ket{z}$ is the position of the cantilever at time $t$.  We now quantize the position function to obtain the effective Hamiltonian acting on the cantilever:
\begin{eqnarray}
\H_C^{'} \: &=& \: \left\{ 
\begin{array}{c c}
		\H_C \, + \, \Gamma \,  \Z^2,  & \: \text{if spin-up, {\it i.e.} $\ket{v_+(t)}$} \\
		\H_C \, - \, \Gamma \, \Z^2,  & \: \text{if spin-down, {\it i.e.} $\ket{v_-(t)}$} \\				
\end{array}
\right. \nonumber \\
&=& \: \H_C \, + \, 2 \, \Gamma \, \Z^2 \, \sigmaZ^{'}
\label{eqn:H_C_eff}
\end{eqnarray}
where we have set
\begin{equation}
\Gamma \: = \: \eta^2/2 \epsilon,
\label{eqn:Gamma}
\end{equation}
and $\sigmaZ^{'}$ is the rotating operator that represents the spin with respect to a reference frame that moves with the effective magnetic field. The total effective Hamiltonian acting on the cantilever $+$ spin system is, thus,
\begin{equation}
\H^{'} \: = \: \H_{C}^{'}.
\label{eqn:tot_H_eff}
\end{equation}

%------------------------------------------------------------------------------------------------------------------------------------------
\section{Modeling decoherence in OSCAR MRFM system}
\label{sec:modeling}
In reality, quantum systems are never completely isolated. The interaction of the system with the rest of the universe is seen as decoherence on the system of interest. There are various decoherence processes that act on both the cantilever and the spin part of the system.

\subsection{Decoherence in the cantilever}
Two primary sources of decoherence act on the cantilever: $(a)$ the thermal environment, and, $(b)$ the continuous measurement of the external read out. Brun and Goan~\cite{Brun03} have shown that these processes can be modeled as Lindblad operators of the form, 
\begin{eqnarray}
\L_k \: &=& \:  A_k \, \Z \: + \:  i \,  B_k \, \p, 
\label{eqn:L1_L2}
\end{eqnarray}
where $\L_1$ captures the effect of thermal environment and  $\L_2$ is due to continuous monitoring of the cantilever position. The coefficients are  
\begin{equation}
\begin{array}{cc}
A_1 \: = \: \sqrt{4 \, \gamma_m \, m \, k_B \, T/\hbar^2}\, , \: &
                         B_1 \: = \: \sqrt{\gamma_m/4 \, m \, k_B \, T}  \\
A_2 \: = \: \sqrt{8 \, \kappa^2 \, E^2/\gamma_c^3}\, , \: & B_2 \: = \: 0.                         
\end{array}
\label{eqn:A1B1A2B2}
\end{equation}
Modeling the thermal environment as a Lindblad operator $\L_1$ requires an additional damping term to be added to the Hamiltonian (refer~\cite{Brun03} Sec.~IV), and the new effective Hamiltonian of the cantilever becomes
\begin{equation}
\H_{eff} \: = \: \H^{'}  \: + \: \gamma_m \, \R,
\label{eqn:H_C_eff_new}
\end{equation}
where $\H^{'}$ is the effective Hamiltonian from Eq.~(\ref{eqn:tot_H_eff}), $\gamma_m$ is the rate of damping due to the thermal environment, and 
\begin{equation}
\R \: \equiv \: \frac{1}{2} (\Z \p \, + \, \p \Z).
\label{eqn:R_defn}
\end{equation}

\subsection{Modeling spin noise}
The source of noise acting on the spin could either be dipole-dipole interactions of the spin with other spins in the lattice, or noise in the magnetic field. The dipole-dipole interaction acts in the laboratory frame while the noise due to magnetic sources can be treated in the reference frame of the effective magnetic field. 

Thermal noise is one of the dominant decoherence process in the MRFM system. This noise affects the cantilever motion, which in turn affects the interaction between the cantilever and the spin. Thus, noise in the cantilever motion means that there is noise in the dipole magnetic field due to cantilever-spin interaction. Also, the external microwave (RF) field is not free of noise. Stipe et al.~\cite{Stipe01} have shown that the spin-lattice relaxation time is quite long, on the order of seconds. This means that spin decoherence is dominated by magnetic noise. As a result, we consider spin relaxation due to magnetic noise and ignore spin-lattice relaxation.

In the reference frame of the effective magnetic field, spin relaxation due to magnetic noise corresponds to simple spin-flip noise. The Lindblad operator corresponding to spin noise can thus be modeled as,
\begin{equation}
\L_3 \: = \: \sqrt{\kappa_s} \, \sigmaX^{'}.
\label{eqn:L3}
\end{equation}
$\kappa_s$ is the rate of spin noise due to magnetic sources, and 
\begin{equation}
\sigmaX^{'} \: = \: \ket{v_{+}(t)} \bra{v_{-}(t)} + \ket{v_{-}(t)} \bra{v_{+}(t)},
\end{equation}
where $\ket{v_{\pm}(t)}$ are the instantaneous eigenstates of the spin Hamiltonian at time $t$ [Eq.~(\ref{eqn:inst_eigenstate})]; note that spin-up or spin-down here corresponds to spin in the direction of, or opposite to, the effective magnetic field. 

For MRFM to be useful as a single spin measurement, it is necessary that the spin decoherence rate $\kappa_s$ be small compared to other parameters in the cantilever-spin system. As we measure the cantilever, we learn something about the state of the spin, and eventually, the spin relaxes to one of its eigenstates, either in the direction of the effective field or in the opposite direction. There are two different time scales at work here: $(i)$ the spin localization time, and, $(ii)$ our observation time; the spin localization time scale is the time it takes, on average, for the spin to collapse to one of its eigenstates due to continuous monitoring of the cantilever; on the other hand, the observation time scale is the time it takes for us to `know' the state of the spin. Typically, the observation time scale is greater than the spin localization time scale as it would take a little longer after the spin is completely relaxed to `know' the state of the spin; however, one can also conclude the state of the spin based on the trend (depending on SNR) in the frequency shift. In the latter case, we do not wait for the spin to localize completely to guess the state of the spin. Irrespective of the method used, the time scale of spin noise ($\kappa_s^{-1}$) has to be longer than both spin localization and observation time scales for MRFM to be useful in single-spin measurement.
 
%\\\\\\\\\\\\\\\\\\\\\\\\\\\\\\\\\\\\\\\\\\\\\\\\\\\\\\\\\\\\\\\\\\\\\\\\\\\\\\\\\\\\\\\\\\\\\\\\\\\\\\\\\\\\\\\\\\\\\\\\\\\\\\\\\\\\\\\\\\\\\\\\\\\\\\\\\\\\\\\\\\\\\\\\\\\\\\\\\\\\\\\\
%
\section{Continuous measurement and feedback in OSCAR protocol}
\label{sec:feedback}
%  Measurement record ---> signal ----> delay ----> feedback
% Discuss how the output from continuous measurement is used to derive feedback to the cantilever

In the OSCAR protocol, the spin orientation is measured by measuring the frequency shift of the cantilever.  A positive gain-controlled feedback mechanism maintains the amplitude of the cantilever at a pre-determined constant, leading to a change in  cantilever frequency.

The continuous monitoring of the cantilever motion is done by optical interferometry. As shown in Fig.~\ref{fig:mrfm_model}, a laser is placed close to the tip of the cantilever; an optical microcavity is formed with the cantilever on one side and the cleaved end of the fiber (laser) on the other. Since the motion of the cantiever is slow compared to the optical frequency, this system can be analyzed in the adiabatic limit. A homodyne measurement is carried out on the light that escapes this cavity. The output of the homodyne measurement corresponds to the position of the cantilever $\expect{\Z}$, and is given by~\cite{Brun03}
\begin{equation}
I_c(t) \: = \: \beta \left( -\, \frac{8 \, e_d \, \kappa \, E}{\gamma_c} \expect{\Z} \: + \: \sqrt{\gamma_c e_d} \: \frac{dW_t}{dt} \right).
\label{eqn:photo_current}
\end{equation}
Here, $I_c(t)$ is the output photocurrent of the homodyne measurement, $\kappa$ is the coupling between the cantilever and the cavity, $\gamma_c$ is the cavity loss rate and $e_d$ is the detector efficiency. Brun and Goan~\cite{Brun03} (Sec.~V) have analyzed this system in detail and we refer to that paper for more information.

The feedback mechanism in the OSCAR protocol is positive gain-controlled, and its objective is to maintain the cantilever amplitude at a pre-determined fixed value. The feedback has the form
\begin{equation}
f(t) \: = \: g \, \times \, (AMP \, - \, Amp(t)) \, \times \, I_c(t \, - \, \Delta),
\label{eqn:fbck}
\end{equation}
where $g$ is the feedback gain, $AMP$ is the pre-determined set point amplitude, $Amp(t)$ is the cantilever amplitude at time $t$, $I_c(t \, - \, \Delta)$ is the delayed output photocurrent, and $\Delta$ corresponds to a delay of $\pi/2$ radians, or equivalently, a fourth of the cantilver oscillation time period. The current amplitude of the cantilever, $Amp(t)$, is derived from the measured output  photocurrent $I_c(t)$, using simple signal processing techniques.

%\\\\\\\\\\\\\\\\\\\\\\\\\\\\\\\\\\\\\\\\\\\\\\\\\\\\\\\\\\\\\\\\\\\\\\\\\\\\\\\\\\\\\\\\\\\\\\\\\\\\\\\\\\\\\\\\\\\\\\\\\\\\\\\\\\\\\\\\\\\\\\\\\\\\\\\\\\\\\\\\\\\\\\\\\\\\\\\\\\\\\\\\
%
\section{Moment equations for OSCAR MRFM with gaussian approximation}
\label{sec:moment_eqns}
% Derive the stochastic moment equation 
In this section, we derive moment equations for the cantilever and
spin. We consider the unitary case first and then move to open system
evolution. We use the Schr\"{o}dinger equation to describe the unitary
evolution, and a full quantum stochastic master equation for the open
system evolution, with continuous monitoring of the cantilever
motion. Finally, we make Gaussian approximation for the cantilever
degree of freedom and show that the moments then obey a closed set of coupled equations. 

\subsection{Unitary evolution}
\label{sec:unitary}
The evolution of the cantilever-spin system is given by a
Schr\"{o}dinger (or a von Neumann) equation:
\begin{equation}
\frac{d\rho}{dt} \: = \: - \left(\frac{i}{\hbar} \right) \, [\H^{'}, \rho]. 
\end{equation}
$\H^{'}$ is the effective Hamiltonian of the cantilever-spin system  [Eq.~(\ref{eqn:tot_H_eff})].

To derive the moment equation for an arbitrary operator $\Op$, we apply the Schr\"{o}dinger equation to obtain equation for $\expect{\Op}$:
\begin{eqnarray}
\frac{d \expect{\Op}}{dt} \: 
&=& \: Tr\left\{ \Op \, \frac{d\rho}{dt} \right\}
= \: \left( \frac{i}{\hbar} \right) \, \expect{[\H^{'}, \Op]}.
\label{eqn:deter_moment_eqn}
\end{eqnarray}

We define (instantaneous) projectors on to the spin-up and spin-down states as:
\begin{equation}
\begin{array}{c}
\hat{\mathcal{P}}_{\uparrow} \: = \: \id \otimes \ket{v_+(t)} \bra{v_+(t)} \\
\hat{\mathcal{P}}_{\downarrow} \: = \: \id \otimes \ket{v_-(t)} \bra{v_-(t)}
\label{eqn:u_d_projectors}
\end{array},
\end{equation} 
where $\id$ acts on the cantilever. The probability of spin-up is defined as
\begin{equation}
r_u \: = \: \expect{\hat{\mathcal{P}}_{\uparrow}}, 
\label{eqn:probup}
\end{equation} 
and, the probability of spin-down is 
\begin{equation}
r_d \: = \: \expect{\hat{\mathcal{P}}_{\downarrow}} \: = \: (1 - r_u).
\label{eqn:probdn}
\end{equation} 
The weight of the spin-up and spin-down wave packets is given by $r_u$ and $r_d$, respectively. Though we can derive the value of $r_d$ from $r_u$ straightforwardly using equation~(\ref{eqn:probdn}), we use $r_d$ to represent spin-down probability for notational convenience. 

To keep track of the cantilever degree of freedom, we define variables that capture the mean position and momenta of the two wave packets as:
\begin{eqnarray}
\left.
\begin{array}{cc}
Z_u \: := \: \expect{\Z \hat{\mathcal{P}}_{\uparrow}}/r_u, & \:
p_u \: := \: \expect{\p \hat{\mathcal{P}}_{\uparrow}}/r_u \\ \\
Z_d \: := \: \expect{\Z \hat{\mathcal{P}}_{\downarrow}}/{r_d}, & \: 
p_d \: := \: \expect{\p \hat{\mathcal{P}}_{\downarrow}}/{r_d} \\
\end{array} 
\right\}.
\label{eqn:wave packets_mean}
\end{eqnarray}

We can now apply Eq.~(\ref{eqn:deter_moment_eqn}) to derive first-order moment equations:
\begin{eqnarray}
{d Z_u}/{dt} \, &=& \,  {p_u}/{m},  \nonumber \\
{d p_u}/{dt} \, &=& \, - (m \omega^2 + 2 \, \Gamma) \, Z_u \: + \: f(t), \nonumber \\
{d Z_d}/{dt} \, &=& \,  {p_d}/{m},  \nonumber \\
{d p_d}/{dt} \, &=& \, - (m \omega^2 - 2 \, \Gamma) \, Z_d \: + \: f(t), \nonumber \\ 
{dr_u}/{dt} \, &=& \, 0.
 \label{eqn:deter_first_order}
\end{eqnarray}
The equations derived above close, which means that higher-order moment terms can be ignored.
The expected position of the cantilever is obtained by the identity
\begin{equation}
\expect{\Z} \: = \: r_u \, Z_u \: + \: r_d \, Z_d.
\label{eqn:expected_pos}
\end{equation}

We observe that the moment equations actually are identical to the equations of motion of a driven (classical) Harmonic oscillator; the resonant frequency of the two wave packets is shifted by an amount $\Delta \omega \approx \Gamma/m \, \omega$, from the natural resonant frequency $\omega$ of the cantilever. We also note that spin-up probability, $r_u$, is a constant of motion. This is consistent with our derivation section~\ref{sec:eff_Hamiltonian}. Spin probabilities, however, are not a constant of motion for the non-unitary evolution considered in the next section. 

\subsection{Open system evolution with continuous monitoring}
\label{sec:open_system}
As discussed in Sec.~{\ref{sec:modeling}, the cantilever-spin system is not isolated, and there are many sources of decoherence that affect both the cantilever and the spin. The sources include thermal noise, noise due to the process of continuous monitoring of the cantilever and noise in the magnetic field. We model these decoherence effects in terms of Lindblad operators. 

There are three Lindblad operators that model the important sources of deocherence, two of which---$\L_1$ and $\L_2$---capture the decoherence acting on the cantilever [Eq.~(\ref{eqn:L1_L2})], while $\L_3$ models the noise on the spin [Eq.~(\ref{eqn:L3})]. $\L_1$ is the noise due to the thermal environment and $\L_2$ is the noise due to continuous monitoring of the cantilever position.

The evolution of the cantilever-spin system is given by a quantum stochastic master equation~\cite{Trajectories}:
\begin{eqnarray}
d \rho \: &=& \: - \left( \frac{i}{\hbar} \right) \, [ \H_{eff}, \rho ] \, dt \: \nonumber \\
&& \, + \, \sum_{k=1}^{3} \, \L_k \, \rho \, \Ldag_k \: - \: \frac{1}{2} \{ \Ldag_k \, \L_k, \rho \} \, dt \nonumber \\
&& \, + \,  \sqrt{e_d} \, \left( (\L_2 - \expect{\L_2}) \, \rho \, - \, \rho \, (\Ldag_2 \, - \, \expect{\Ldag_2}) \, \right) \: dW_t , 
\nonumber \\
\label{eqn:mrfmsme}
\end{eqnarray}
where $e_d$ is the detector efficiency; $dW_t$ is a stochastic (white) noise process with the property: $M[dW_t] \: = \: 0$ and  $dW_t^2 = dt$; and $\H_{eff}$ is the effective Hamiltonian acting on the system [Eq.~(\ref{eqn:H_C_eff_new})].

As in the unitary case, we consider an arbitrary operator $\Op$ acting on the system (with evolution described in Eq.~(\ref{eqn:mrfmsme})), and derive its  moment equations. Using Eq.~(\ref{eqn:mrfmsme}), the first-order moment equation for $\expect{\Op} \, = \, Tr\{\Op \rho\}$ is
\begin{eqnarray}
  d \expect{\Op} \, &=& \,  \frac{i}{\hbar} \expect{[\H_{eff}, \Op]} \, dt  \nonumber \\
&& \; + \,  \sum_{k=1}^3 \, \expect{\Ldag_k \Op \L_k}  - \frac{1}{2} \expect{\{ \Ldag_k \L_k, \Op \} } \, dt 
			\nonumber \\
&& \: + \:  \sqrt{e_d} \, \left( \expect{\Op \L_2} \: - \: \expect{\Op} \expect{\L_2}  \: \right. \nonumber \\
&& \;\;\;\;\;\;\;\;\;\;\;\;+ \:  \left. \expect{\Ldag_2 \Op} \: - \:  \expect{\Ldag_2} \expect{\Op} \right) \, dW_t. 
\label{eqn:momenteqn}
\end{eqnarray}

We describe the (mean) position and momentum of the cantilever in terms of normalized wave packets with spin up or down, respectively [Eq.~(\ref{eqn:wave packets_mean})]. They are of the form: 
\begin{equation}
\expect{\Op}_u =  \expect{\Op \hat{\mathcal{P}}_{\uparrow}}/r_u \: \: \text{and} \:  \: \expect{\Op}_d = \expect{\Op \hat{\mathcal{P}}_{\downarrow}}/r_d, 
\label{eqn:Op_u_Op_d}
\end{equation}
where  $r_u$ and $r_d$ are the spin probabilities as defined in equations~(\ref{eqn:probup}) and (\ref{eqn:probdn}). As in the unitary case, the mean $\expect{\Op}$ is defined as
$$
\expect{\Op} \, = \, r_u \, \expect{\Op}_u \, + \, r_d \, \expect{\Op}_d.
$$  
The system evolution described in Eq.~(\ref{eqn:mrfmsme}) is stochastic. This means that the moments like $\expect{\Op}_u$ and $\expect{\Op}_d$ evolve stochastically as well; the moment equation (\ref{eqn:momenteqn}) is a stochastic differential equation,  unlike its unitary counterpart Eq.~(\ref{eqn:deter_moment_eqn}). These equations follow stochastic calculus described by It$\hat{\text{o}}$ rules~\cite{Oksendal03}.  To give the evolution of  $\expect{\Op}_u$ and $\expect{\Op}_d$,  we apply It$\hat{\text{o}}$ rules and get,
\begin{eqnarray}
d  \expect{\Op}_u \: 
&=& \: d \expect{\Op}/r_u \: - \: \expect{\Op}\, dr/{r_u^2} \: - \:  d \expect{\Op} \, dr/r_u^2  \nonumber \\
			&& \;\; + \: \expect{\Op} (dr)^2/{r_u^3},
\label{eqn:op_by_r} 
\end{eqnarray}
and similarly for $\expect{\Op}_d$.

%---------------------------------------------------------------------------------------------------------------------
%
\subsubsection{First-order moment equations}
\label{sec:first_order}
We now get to the task of deriving moment equations for the cantilever and the spin. As in the unitary case, we first derive the equation for spin-up probability $r_u = \expect{\hat{\mathcal{P}}_{\uparrow}}$, before we derive equations for the wave packets that describe the cantilever motion. 

Using Eq.~(\ref{eqn:momenteqn}), we derive the equation for spin-up probability
\begin{eqnarray}
dr_u \: &=& \: \kappa_s \, (1 \, - \, 2 r_u) \, dt  \nonumber \\
&& \; + \, 2 \, \sqrt{e_d} \, A_2 \, r_u \, r_d \, (Z_u \, - \, Z_d)\, dW_t. 
\label{eqn:sme_r}
\end{eqnarray}
Here, $Z_u$ and $Z_d$ are mean positions of the cantilever wave packets for spin up or down, respectively; $\kappa_s$ is the rate of spin noise [Eq.~(\ref{eqn:L3})]; and $A_2$ is part of the Lindblad operator for cantilever decoherence due to continuous measurement [Eqs.~(\ref{eqn:L1_L2}) and (\ref{eqn:A1B1A2B2})]. We used the identity in Eq.~(\ref{eqn:expected_pos}) in our derivation. Note that unlike the unitary case, Eq.~(\ref{eqn:deter_first_order}), the spin-up probability $r_u$ is not a constant of motion. However, since  $r_d = (1-r_u)$, the single variable $r_u$ is enough to capture the evolution of spin probabilities.

The motion of the cantilever is described in terms of two wave packets corresponding to the spin being in the up or down state, respectively. Each wave packet is described by its position and momentum. To derive the moment equation for these variables, we apply Eqs.~(\ref{eqn:op_by_r}) and (\ref{eqn:momenteqn}), to obtain
\begin{eqnarray}
d Z_u \: &=& \:  \left( (p_u/m) \, 
				- \, 4 \, e_d \, A_2^2 \, \DZ_u \, r_d \, (Z_u \, - \, Z_d) \right) \, dt \nonumber \\
			&& \: - \: \kappa_s \, (r_d/r_u)  \, (Z_u \, - \, Z_d) \, dt  \nonumber \\
			&& + \: 2 \, \sqrt{e_d} \, A_2 \, \DZ_u \, dW_t , 
\label{eqn:Zu}
\end{eqnarray}
and
\begin{eqnarray}
 d p_u \: &=& \: \left(\, - \, (m \omega^2 + 2 \, \Gamma ) \, Z_u \,  
 				- \: 2 \gamma \, p_u  \,  + \, f(t) \, \right) dt  \nonumber \\
          		&& - \, \kappa_s ( r_d/r_u )\, (p_u \, - \, p_d) \, dt \,  \nonumber \\ 
 	     && - \: 4 \, e_d \, A_2^2 \, \DR_u \, r_d \, (Z_u \, - \, Z_d) \, dt \: \nonumber \\
	     && + \: 2 \, \sqrt{e_d} \, A_2 \, \DR_u \, dW_t.   
\label{eqn:pu}
\end{eqnarray}
Similarly,
\begin{eqnarray}
d Z_d \: &=& \:  \left( (p_d/m) \: + \: 4 \, e_d \, A_2^2 \, \DZ_d \, r_u \, (Z_u \, - \, Z_d) \right) \, dt \nonumber \\
			&& + \: \kappa_s \, ( r_u/r_d ) \, (Z_u \, - \, Z_d) \, dt \nonumber \\
                        && + \: 2 \, \sqrt{e_d} \, A_2 \, \DZ_d \, dW_t ,
 \label{eqn:Zd}
 \end{eqnarray}
 and
 \begin{eqnarray}
d p_d  \: &=&\:  \left( - \,(m \omega^2 - 2 \, \Gamma )  \, Z_d  \: - \: 2 \gamma \, p_d  
  				\: + \: f(t) \right) \, dt \nonumber \\
		&&  + \, \kappa_s  (r_u/r_d )\, (p_u \, - \, p_d) \, dt   \nonumber \\
		&&  + \:  4 \, e_d \, A_2^2 \, \DR_d \, r_u \, (Z_u \, - \, Z_d) \, dt \nonumber \\
		&&  + \: 2 \, \sqrt{e_d} \, A_2 \, \DR_d \, dW_t.  
\label{eqn:pd}			
\end{eqnarray}

In the unitary case, the first-order moment equations (\ref{eqn:deter_first_order}) formed a closed set. In the open system evolution however, the first order moment equations~(\ref{eqn:Zu}$-$\ref{eqn:pd}) do not form a closed set, and depend on second-order terms like $\DZ_u$ and $\DR_u$:
\begin{eqnarray}
\DZ_u &:=&  \left( \expect{\Z^2}_u \, - \, Z_u^2 \right), \label{eqn:DZ_u_def} \\ 
\DR_u &:=&  \left( \expect{\R}_u \, - \, Z_u p_u \right),  \label{eqn:DR_u_def} \\
\DZ_d &:=& \left( \expect{\Z^2}_d \, - \, Z_d^2  \label{eqn:DZ_d_def}  \right),\\
\DR_d &:=& \left( \expect{\R}_d \, - \, Z_d p_d \label{eqn:DR_d_def} \right),
\end{eqnarray}
where $\R$ was defined in (\ref{eqn:R_defn}) above.  Since the set of first-order equations do not close, we need to include second-order terms, and derive equations for them as well.
 
 %------------------------------------------------------------------------------------------------------------------------------------------
%
\subsubsection{Second-order moment equations and Gaussian approximation}
\label{sec:second_order}
We will need equations for six second-order equations: $\DZ_u$, $\Dp_u$, $\DR_u$, $\DZ_d$, $\Dp_d$ and $\DR_d$.   We begin with the derivation of $\DZ_u$. 

To derive the moment equation, we first expand $\DZ_u$ in terms of its definition~(\ref{eqn:DZ_u_def}), then apply It$\hat{\text{o}}$ rules where necessary as shown in Eq.~(\ref{eqn:op_by_r}) for an arbitrary operator. Finally, we expand using the moment equation~(\ref{eqn:momenteqn}) to get:
\begin{eqnarray}
&&d \DZ_u \, = \,  \left( 2  \DR_u/m \, +  \, B_1^2 \hbar^2 \right) \, dt \nonumber \\
&& \; \; - \, 4 \, e_d \, A_2^2  \left(  \DZ_u \, + \,  r_d \, (Z_u \, - \, Z_d) \, \DZcube_u  \right) \, dt \nonumber \\
&& \; \;  - \, \kappa_s \, ( r_d/r_u ) \, \left( \DZ_u \, - \, \DZ_d \, \right. \nonumber \\
&& \; \;\; \;\; \; \; \;\; \;\; \;\; \;\; \;\; \;\; \;\; \;\; \; \; \;\; \;\; \;    - \, \left. (Z_u \, - \, Z_d)^2 \frac{}{} \right) \, dt \, \nonumber \\
&& \: \: + \, 2 \, \sqrt{e_d} \, A_2 \DZcube_u \, dW_t. 
\label{eqn:DZ_u}
\end{eqnarray}
We have used some of the first order equations in the above derivation. 

We derive the equation for $\DR_u$, defined in Eq.~(\ref{eqn:DR_u_def}), by the same procedure used to derive Eq.~(\ref{eqn:DZ_u}), and obtain: 
\begin{eqnarray}
&&d\DR_u \: = \,  - \, ( m\omega^2  \, + 2 \, \Gamma ) \, \DZ_u \, dt \nonumber \\
&&\; \; + \, \left( \Dp_u/m \, - \, 2 \, \gamma \,  \DR_u \right) \, dt \nonumber \\ 
&& \;\; -\, 4 \, e_d \, A_2^2  \left( \DR_u  \DZ_u \, \right.  \nonumber \\
&&  \; \;\; \;\; \; \; \;\; \;\; \;\; \;\; \;\; \;\;\;  +  \, \left.  r_d \, (Z_u - Z_d) \, \DS_u \right) \, dt \nonumber \\
&&\;\;  - \, \kappa_s ( r_d/r_u ) \left( \DR_u  -  \DR_d \, \right. \nonumber \\
&& \; \; \; \;\;\; \;\; \;\; \;\; \;\;\; \;\; \;\; \;\; \;\; \;	- \, \left. (Z_u  -  Z_d) (p_u  -  p_d)  \frac{}{} \right) dt \nonumber \\	
&& \;\;  + \: \sqrt{e_d} \, A_2 \, \DS_u \, dW_t.	
\label{eqn:DR_u}
\end{eqnarray}

Note that a new second order term $\Dp_u$ appears in the above equation, and is defined as
\begin{eqnarray}
\Dp_u &:=&  \left( \expect{\p^2}_u \, - \, p_u^2 \right).
\label{eqn:DP_u}
\end{eqnarray}
Following the same procedure, we construct the moment equation for $\Dp_u$:
\begin{eqnarray}
&&d \Dp_u \, = \,  \Sigma_{k=1}^2 \, A_k^2\, \hbar^2  \, dt  \nonumber \\
&& \; \;  - \, 4 \, \left( ( m \omega^2/2 \, + \, \Gamma ) \, \DR_u  + \gamma \Dp_u \right) \, dt \nonumber \\
&& \;\; - \,  2 \, e_d \, A_2^2  \left(  2 \, \DR_u^2  \, \right. \nonumber \\
&& \;\;\;\;\;\;\;\; \;\;\;\;\;\;\;\;   + \, \left. r_d \, (Z_u - Z_d) \, \DQ_u \right) \, dt \nonumber \\
&& \; \;  - \, \kappa_s \,(r_d/r_u ) \left( \Dp_u \, - \, \Dp_d \, \right. \nonumber \\
&& \;\;\;\;\;\;\;\; \;\;\;\;\;\;\;\;\;\;\;\;\;\;\;\;\;\; \;\;\;\;\;\;\;\;\;\;    - \, \left. (p_u \, - \, p_d)^2 \right) \, dt \nonumber \\  
&& \; \; + \, \sqrt{e_d} \, A_2 \, \DQ_u \, dW_t,
\label{eqn:Dp_u}
\end{eqnarray}

Similarly, we derive equations for the second-order moment terms of the spin-down wave packet:
\begin{eqnarray}
&& d \DZ_d \: = \, \left(2  \DR_d/m \, + \,  B_1^2 \hbar^2  \right) \, dt \nonumber \\
&& \;\;  - \, 4 \, e_d \, A_2^2 \left( \DZ_d \, -  \, r_u \, (Z_u \, - \, Z_d) \, \DZcube_d \right) \, dt \, \nonumber \\ 
&& \; \;  + \, \kappa_s \, ( r_u/r_d ) \left( \DZ_u  -  \DZ_d \, \right. \nonumber \\ 
&& \;\;\;\;\;\;\;\;\;\;\;\;\;\;\;\;\;\;\;\;\;\;\;\;\;\;\;\;\;\;\;\;	+ \, \left. (Z_u  -  Z_d)^2 \frac{}{} \right)\, dt \, \nonumber \\
&& \;\; + \, 2 \, \sqrt{e_d} \, A_2 \DZcube_d \, dW_t,  
\label{eqn:DZ_d} 
\end{eqnarray}
\begin{eqnarray}
&&d \DR_d \: = \, - \, \left( m\omega^2  \, - \, 2 \, \Gamma \right) \DZ_d \, dt \nonumber \\
&& \; \;  + \, \left( \Dp_d/m \, - \, 2 \, \gamma \, \DR_d \right) \, dt  \nonumber \\
&& \;\; -  \,  4 \, e_d \, A_2^2 \left( \DR_d \, \DZ_d \, \right. \nonumber \\
&& \;\;\;\;\;\;\;\; \;\;\;\;\;\;\;\;     -  \left. \, r_u \, (Z_u \, - \, Z_d) \, \DS_d \right) \, dt \nonumber \\
&& \;\; + \, \kappa_s ( r_u/r_d ) \left( \DR_u -  \DR_d  \right. \nonumber \\
&& \;\;\;\;\;\;\;\; \;\;\;\;\;\;\;\; \;\;\;\; \;\;\;\; \;\;\;\;     + \left. (Z_u -  Z_d)\, (p_u  -  p_d)  \right) \, dt \nonumber \\	
&& \;\; + \, \sqrt{e_d} \, A_2 \, \DS_d \, dW_t,	
\label{eqn:DR_d} 	 
\end{eqnarray}
\begin{eqnarray}
&&d \Dp_d \: = \,  \Sigma_{k=1}^2 \, A_k^2\,  \hbar^2 \, dt \nonumber \\
&& \; \; - \, 4 \, \left( \,(m \omega^2/2 \, - \,  \Gamma) \, \DR_d \, + \, \gamma \, \Dp_d \right) \, dt \nonumber \\
&& \;\; - \, 2 \, e_d \, A_2^2 \, \left( 2 \, \DR_d^2 \, \right. \nonumber \\
&& \;\;\;\;\;\;\;\;\;\;\;\;\;\;\;\;  - \, \left. r_u \, (Z_u \, - \, Z_d) \, \DQ_d  \right) \, dt \nonumber \\
&&\;\;  + \, \kappa_s \, (r_u/r_d) \left( \frac{}{} \Dp_u  -  \Dp_d \, \right. \nonumber \\
&&\;\;\;\;\;\;\;\; \;\;\;\;\;\;\;\; \;\;\;\; \;\;\;\; \;\;\;\; \;\;\;\; \;\;\;\;  + \, \left. (p_u  -  p_d)^2 \frac{}{} \right) \, dt \nonumber \\  
&& \;\; + \, \sqrt{e_d} \, A_2 \, \DQ_d \, dW_t.  
\label{eqn:Dp_d}
\end{eqnarray}

We have now derived equations for all the second order terms that appear in the equations. However, there are new terms, of third order, that appear in the second order equations~(\ref{eqn:DZ_u}$-$\ref{eqn:Dp_d}):
 \[
\DZcube_{u,d}, \,  \DQ_{u,d} \: \text{and} \: \DS_{u,d}. 
\]
The second order equations~(\ref{eqn:DZ_u}$-$\ref{eqn:Dp_d}) do not close as they depend on terms of third order. We now make the Gaussian approximation for the cantilever. Assuming that the wave packets were initially Gaussian, and that they remain approximately Gaussian at later times, then the third order terms vanish. The system of first-order equations along with the second order equations now close, giving us a set of $11$ coupled equations. 

We will now verify the validity of this approximation by numerically comparing the approximate solution to that of the full quantum stochastic master equation~(\ref{eqn:mrfmsme}). 

%\\\\\\\\\\\\\\\\\\\\\\\\\\\\\\\\\\\\\\\\\\\\\\\\\\\\\\\\\\\\\\\\\\\\\\\\\\\\\\\\\\\\\\\\\\\\\\\\\\\\\\\\\\\\\\\\\\\\\\\\\\\\\\\\\\\\\\\\\\\\\\\\\\\\\\\\\\\\\\\\\\\\\\\\\\\\\\\\\\\\\\\\
%
\section{Numerical results}
\label{sec:results}
We simulated the full quantum stochastic master equation using a $C++$  quantum master equation library developed by Brun  and Shaw~\cite{SME-Package}. We use a fifth order Runge-Kutta integrator~\cite{Press07} to simulate the set of coupled equations~(\ref{eqn:sme_r}$-$\ref{eqn:pd}) and (\ref{eqn:DZ_u}$-$\ref{eqn:Dp_d}) for the Gaussian approximation. 

We choose parameters based on those used by Berman et al.~\cite{Berman04} and Brun and Goan~\cite{Brun03, BrunGoan05}. The values of parameters, in dimensionless units, are:
\begin{eqnarray}
\begin{array}{c}
\hbar = \omega = m = 1  \\
 \eta \, = \, 0.6, \: \epsilon \, = \, 100, \: \gamma \, = \, \omega/Q \, = \, 10^{-5},  \: e_d \, = \, 0.85 \nonumber \\
 A_1 \, = \, 0.2, \: B_1 \, = \, 5 \times 10^{-5}, \: A_2 \, = \, 0.07, \:  B_2 \, = \, 0. 
\end{array} \\
\label{eqn:params1}
\end{eqnarray}
We assume the physical units in which $\omega = 10^5 \, \text{s}^{-1}$, $m = 10^{-12}$ kg, $\eta =  3\times 10^7$ T/m, and $\epsilon = 300 \, \mu$T, consistent with current experiments. The values of $A_1$ and $B_1$ are different from the ones used in \cite{Brun03}. We assume that, in dimensionless units,
\[
k_B T \: = \: 10^3,
\]
in our present calculations. This value of $k_BT$ corresponds to a temperature of about $10$ mK; though this is lower than the temperature accessible currently (which is around  $300$ mK~\cite{Degen09}), given the considerable progress made on the experimental front in the last few years, the temperature we have used should be accessible in the near future.

% Re-scaling the measurement
The output of the homodyne measurement that measures the cantilever position is given by Eq.~(\ref{eqn:photo_current}), and can be expressed (in rescaled units) as
\begin{equation}
I_c(t)\, dt \: = \:  \expect{\Z} \, dt \: - \: \Lambda \,  \: dW_t ,
\label{eqn:photo_current_rescaled}
\end{equation}
where
\begin{equation}
\Lambda \: = \: \frac{50}{8 \, \kappa \, E \, Q} \,  \sqrt{\frac{\gamma_c^3}{e_d}} .
\label{eqn:Lambda}
\end{equation}
Here, $Q$ is the quality factor of the cantilever, and the other parameters are due to the cantilever-light interaction, which is part of the optical interferometry process (see Sec.~\ref{sec:feedback}). Since the driving force applied to the cantilever is part of a feedback process, the driving frequency  is resonant with the cantilever frequency. We know that, in the resonant case, the steady state amplitude of a Harmonic oscillator is proportional to its quality factor~\cite{Albrecht91}. In our simulation, we assume that the cantilever is in its steady state before it interacts with the spin, and that the steady state amplitude is, in dimensionless units, $50$, which corresponds to $32$ nm in physical units (in \cite{Rugar04}, the peak amplitude is $16$ nm and the quality factor is $50,000$); due to this rescaling of amplitude, the cantilever quality factor $Q$ appears in the above equation~(\ref{eqn:Lambda}), and we set $\beta = - 50/Q$ in Eq.~(\ref{eqn:photo_current}). 

We can find the value of the coefficient $\Lambda$ straightforwardly, by noting that (in dimensionless units)
\[
\frac{8 \kappa E}{\gamma_c} \: = \: 1.9 \times 10^3,
\]
and
\[
\gamma_c \: = \: 1.4 \times 10^{8}.
\]
See \cite{Brun03} (Sec.~VII) for more information.   

The positive gain controlled feedback function $f(t)$ is defined in  Eq.~(\ref{eqn:fbck}), and depends on the feedback gain $g$ and the pre-determined set point amplitude $AMP$. In our simulation, we set these parameters to
\begin{eqnarray}
g = 0.0001 \: \: \text{and} \: \: AMP = 50.
\label{eqn:params2}
\end{eqnarray}
We chose a small value for $g$ to avoid numerical instability when simulating the fully quantum stochastic master equation. However, the accuracy of the approximation is unaffected by larger values of $g$.

\subsection{Accuracy of Gaussian approximation}
\label{sec:results-gaus}
We now numerically demonstrate the accuracy of the Gaussian approximation by comparing it to a fully quantum stochastic master equation~(\ref{eqn:mrfmsme}). The parameters used to generate these plots are given in equations~(\ref{eqn:params1}) and (\ref{eqn:params2}); the spin noise rate, $\kappa_s$, is set to $0.001$. We assume that the cantilever and spin are initially decoupled; we also assume that the spin is in an equal superposition of up and down state, while the cantilever is at its lowest position ($-AMP$).

In Figure~\ref{fig:sme_vs_gaus}, we plot the time evolution of the cantilever position and that of the spin-up probability; Figs.~\ref{fig:expect_pos_1}, \ref{fig:expect_pos_2} and \ref{fig:expect_pos_3}, shows the time evolution of the expected position of the cantilever $\expect{\Z}$, as defined in Eq.~(\ref{eqn:expected_pos}), at different times, while Fig.~\ref{fig:spin_prob} plots the time evolution of the spin-up probability $r_u = \expect{\hat{\mathcal{P}}_{\uparrow}}$. Figure~\ref{fig:sme_vs_gaus} shows that the Gaussian approximation tracks the quantum stochastic master equation very closely. 

The spin probability does not relax completely (to either $0$ or $1$); after initial fluctuations, the spin probability gets closer to $1$ (spin-up), the spin `flips' and the probability is then closer to zero (spin-down), and finally, another `flip' takes it back closer to $1$ (spin-up). The spin exhibits this behavior because the value of $\kappa_s$ is relatively high ($= 0.001$), leading to multiple `flips' in spin probability. 
\begin{figure*}[htp]
   \begin{center}
   \begin{tabular}{cc}
	\subfigure[]{\label{fig:expect_pos_1} \includegraphics[height=2.5in]{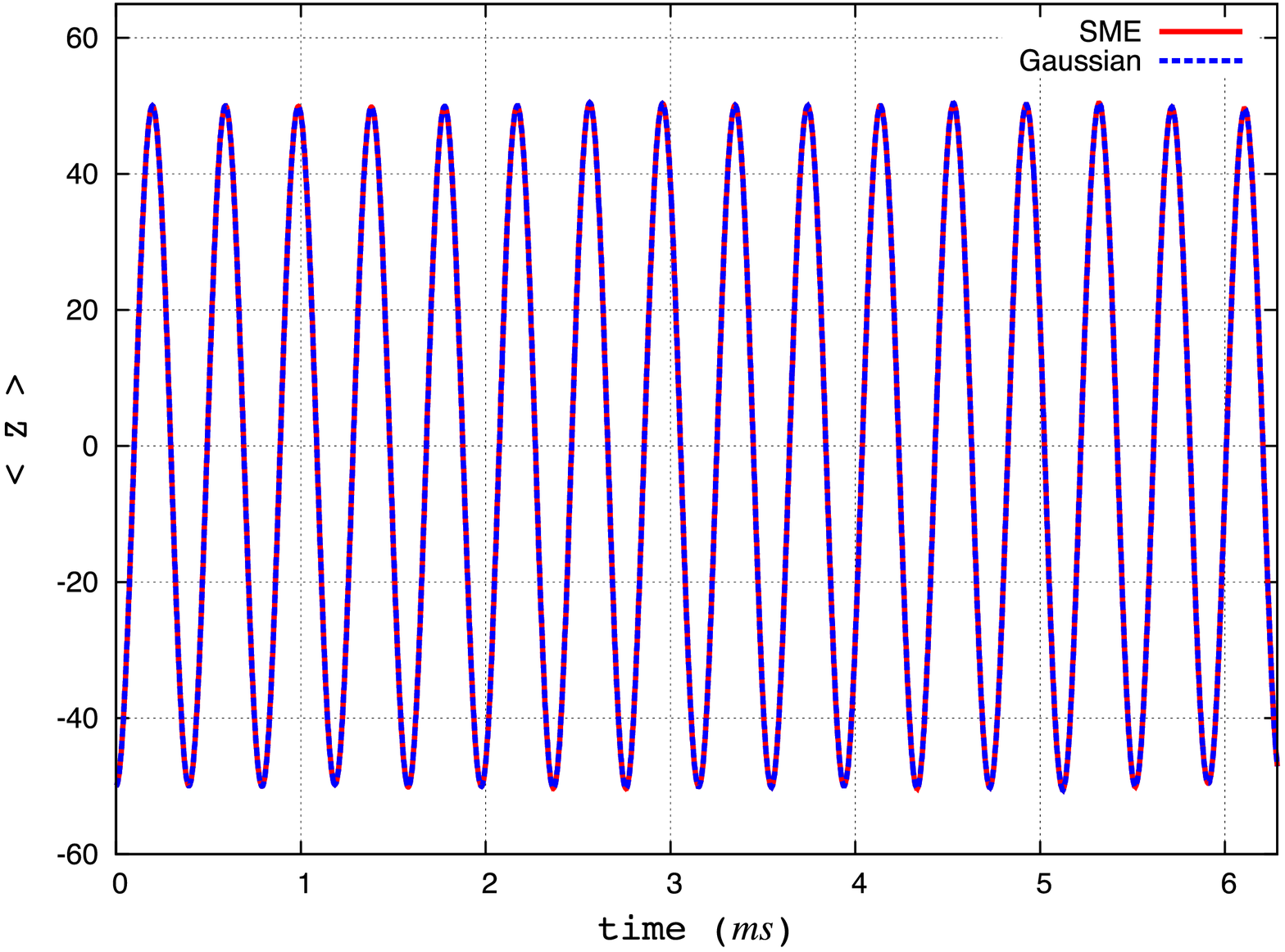}} &
	\subfigure[]{\label{fig:expect_pos_2} \includegraphics[,height=2.5in]{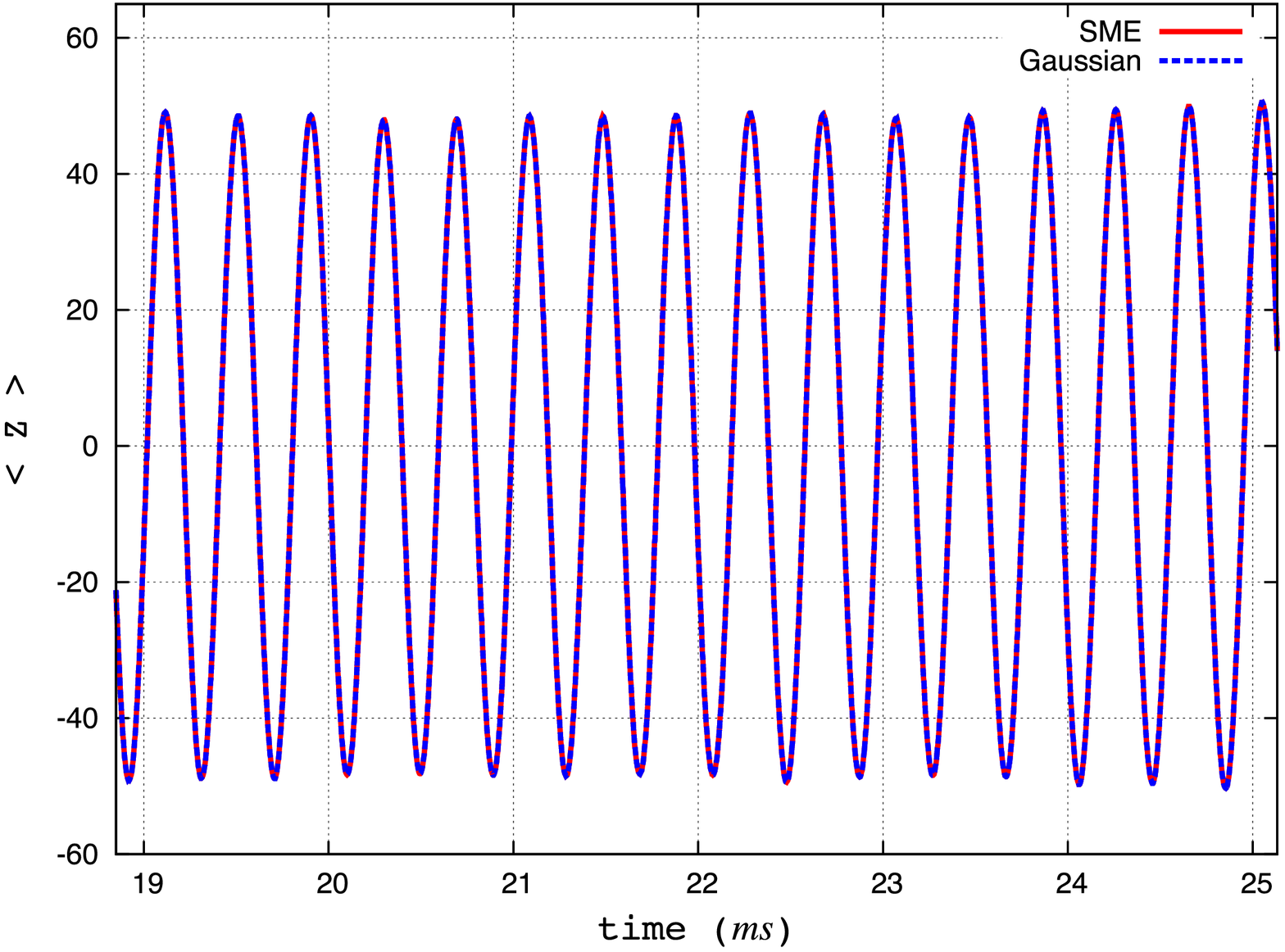}} \\
	\subfigure[]{\label{fig:expect_pos_3} \includegraphics[height=2.5in]{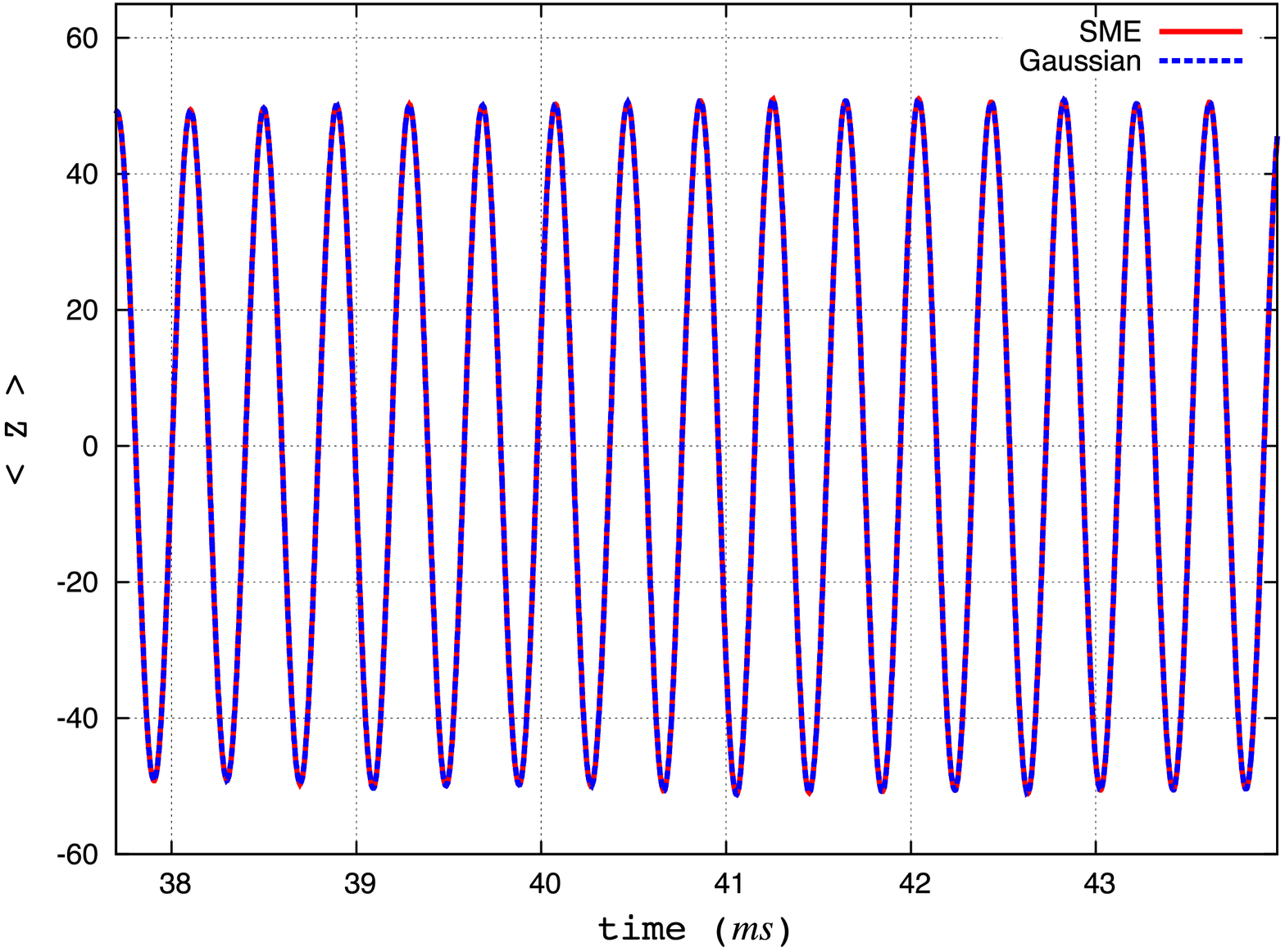}} &
	\subfigure[]{\label{fig:spin_prob} \includegraphics[height=2.5in]{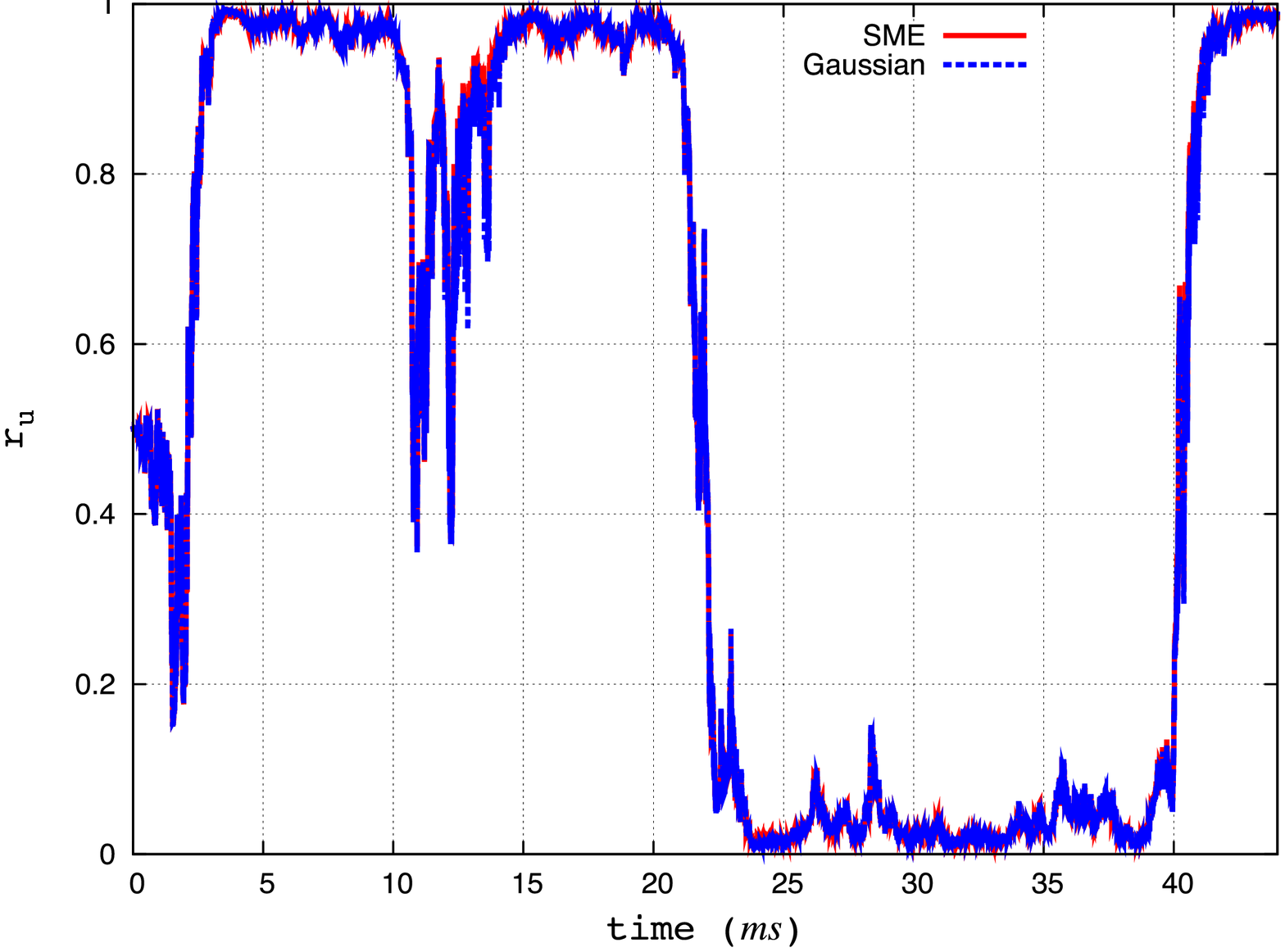}}\\ 
  \end{tabular}	
  \end{center}
  \caption{(Color online.) Comparing full quantum SME and Gaussian
    approximation. We plot $\expect{\Z}$ as a function of time for both  the quantum SME as well and the Gaussian approximation. The parameter values used in this simulation are defined in equations~(\ref{eqn:params1}) and (\ref{eqn:params2}); the rate of spin noise $\kappa_s$ is set to $0.001$. We assume that the cantilever-spin system is initially decoupled, with the cantilever at the lowest point in its oscillation, while the spin is in an equal superposition of up and down states. In Figs.~\ref{fig:expect_pos_1}, \ref{fig:expect_pos_2} and \ref{fig:expect_pos_3}, we plot the expected position of the cantilever $\expect{\Z}$ at different times, and in Fig.~\ref{fig:spin_prob}, we plot the evolution of spin-up probability $r_u = \expect{\hat{\mathcal{P}}_{\uparrow}}$.}
  \label{fig:sme_vs_gaus}
\end{figure*}

\subsection{OSCAR MRFM and single-spin measurement}
\label{sec:results-single_spin}
%--- 3 different values of kappa
We now address the question of how high $\kappa_s$ can be, for OSCAR MRFM to be a useful single-spin measurement device. Henceforth we use the numerically efficient Gaussian approximation equations to describe the OSCAR MRFM system. 

In Figure~\ref{fig:3kappas}, we plot the time evolution of spin-up probability for different values of $\kappa_s$: ($i$) $10^{-3}$, ($ii$) $10^{-4}$ and $10^{-5}$ in dimensionless units ($16$ Hz, $1.6$ Hz and $160$ mHz, respectively, in physical units); the parameters and the initial conditions are the same as in Fig.~\ref{fig:sme_vs_gaus}. We observe that the number of spin flips decreases with the value of $\kappa_s$. Intuitively, one can see from Fig.~\ref{fig:3kappas} that smaller values of $\kappa_s$ are preferable, as we must track the cantilever position for a sufficient length of time to determine the orientation of the spin with required accuracy. If $\kappa_s$ is higher, the likelihood of a  spin flip increases, corrupting the samples used to determine the spin orientation. 
\begin{figure*}[htp]
  \begin{center}
    	 \includegraphics[width = 4.0in, height=3.75in]{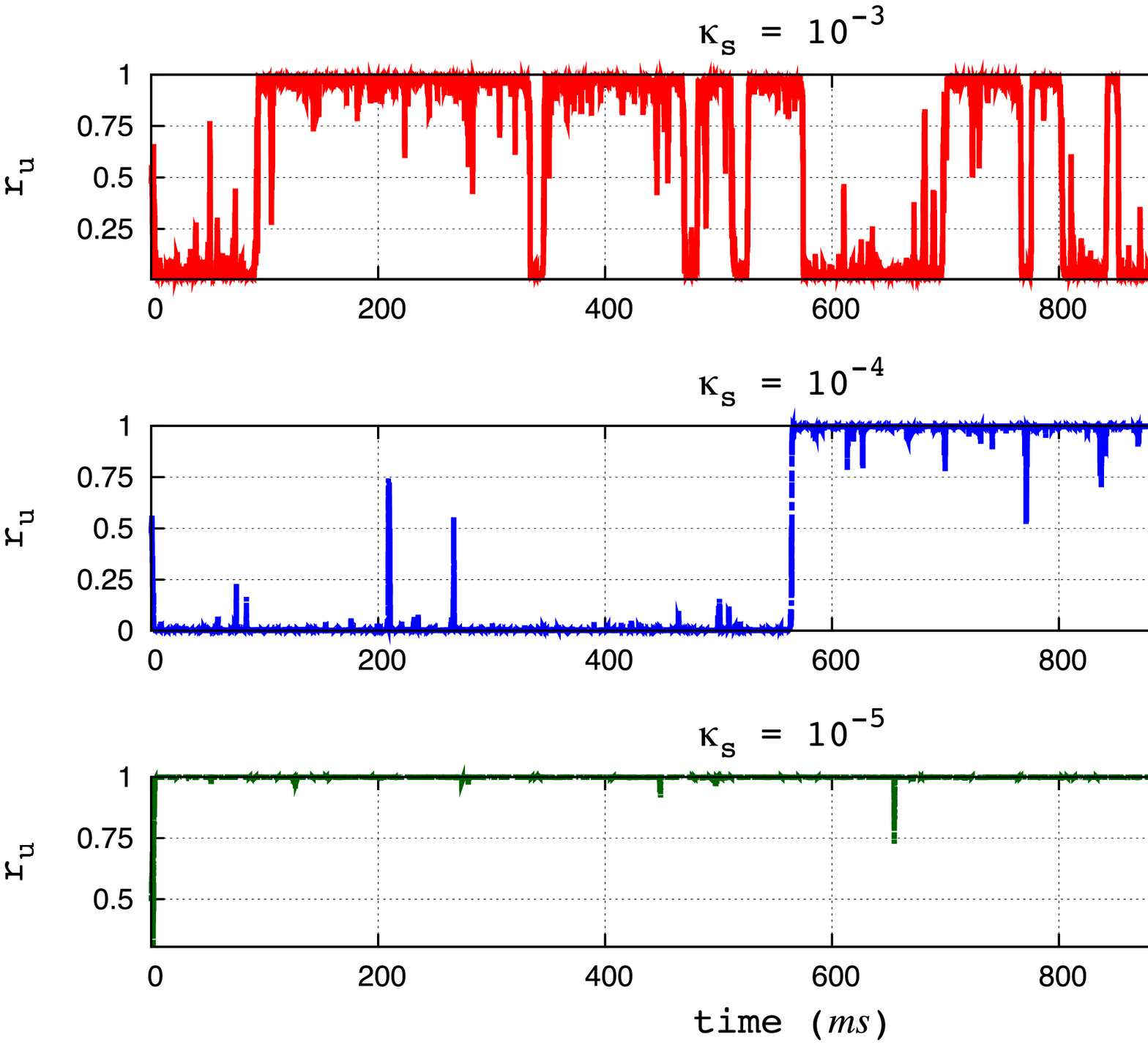}
  \end{center}
  \caption{(Color online.) Spin probability for different $\kappa_s$. We plot the time evolution of spin-up probability, $r_u = \expect{\hat{\mathcal{P}}_{\uparrow}}$, for $3$ different values of spin noise rate $\kappa_s$: ($i$) $10^{-3}$,  ($ii$) $10^{-4}$, and ($iii$) $10^{-5}$. The number of spin flips decreases as $\kappa_s$ is reduced.}
  \label{fig:3kappas}
\end{figure*}

In the OSCAR protocol, we measure the spin by measuring the resonant frequency of the cantilever; more precisely, the shift in the cantilever frequency. To determine the cantilever frequency numerically, we use the Fast Fourier Transform (FFT) algorithm {\it dfour1} from~\cite{Press07}. The (numerical) frequency resolution of this FFT is
\begin{eqnarray}
\Delta f \approx \frac{1}{N \, \Delta t} \approx \frac{1}{T_{Sampling}},  \label{eqn:freq_res}
\\ \nonumber
\end{eqnarray}
where $N$ is the number of samples, $\Delta t$ is the sample spacing, and $\Delta f$ is the frequency resolution. $T_{Sampling}$ is the total sampling time. 

We assume that the sample spacing $\Delta t$ is fixed. The shift in the resonant frequency is proportional to $\Gamma$ [Eq.~(\ref{eqn:Gamma})]. For the parameters chosen in our simulation [Eq.~(\ref{eqn:params1})], the value of $\Gamma$ is $1.8\times 10^{-3}$. Thus, the frequency resolution to be achieved should satisfy 
\begin{equation}
\Delta f \le \: \Gamma/2\pi \approx \: 2.8\times 10^{-4}. \label{eqn:bound_freq_res}
\end{equation} 
The sample spacing used in our simulation is $\Delta t = 0.02$; setting the frequency resolution $\Delta f = 10^{-4}$, and using Eq.~(\ref{eqn:freq_res}), one can easily derive $N=5\times 10^5$. It is numerically convenient to the FFT algorithm if $N$ is a power of $2$; we thus choose $N = 2^{19}$ (the power of $2$ closest to $5\times10^5$) for our simulation, and the frequency resolution becomes $\Delta f = 9.5 \times 10^{-5}$. 

%--- 2 different trajectories for kappa = 10^-5 and spin relaxation time
%--- calculation of frequency using FFT
Given the sample spacing, $\Delta t$, and the number of samples, $N$, the total sampling period in dimensionless (time) units is 
\begin{equation}
T_{Sampling} = N \Delta t \approx 10,500. \label{eqn:sample_period}
\end{equation}
The time scale of spin noise, $\kappa_s^{-1}$, must be longer than the sampling period $T_{Sampling}$, for OSCAR MRFM system to be useful in single-spin measurement. Thus, we set $\kappa_s = 10^{-5}$ in our simulation.

In Figure~\ref{fig:oscar_freqShift} we show the frequency shift achieved in the OSCAR protocol. We plot two different trajectories: one in which the spin relaxes to the spin-up state, and another in which it relaxes to the spin-down state. Figure~\ref{fig:spinprob_2trajs} plots the time evolution of the spin-up probability $r_u$ for the two trajectories, and Fig.~\ref{fig:freqShift} displays their corresponding Fourier amplitudes. It is clear from Fig.~\ref{fig:freqShift} that there is a marked shift in resonant frequency: to the right in the case of spin-up, and to the left in the case of spin-down; we also observe that this shift is proportional to $\Gamma$ ($=1.8 \times 10^{-3}$). (Note that the natural resonant frequency of the cantilever, in dimensionless units, is $\omega = 1$.) A shift to the right in the  resonant frequency of the cantilever implies that the spin is in the spin-up state, while a shift to the left indicates that the spin is in the spin-down state.
\begin{figure*}[htp]
  \begin{center}
      \begin{tabular}{cc}
    	 \subfigure[]{\label{fig:spinprob_2trajs} \includegraphics[width = 3.0in, height=2.5in]{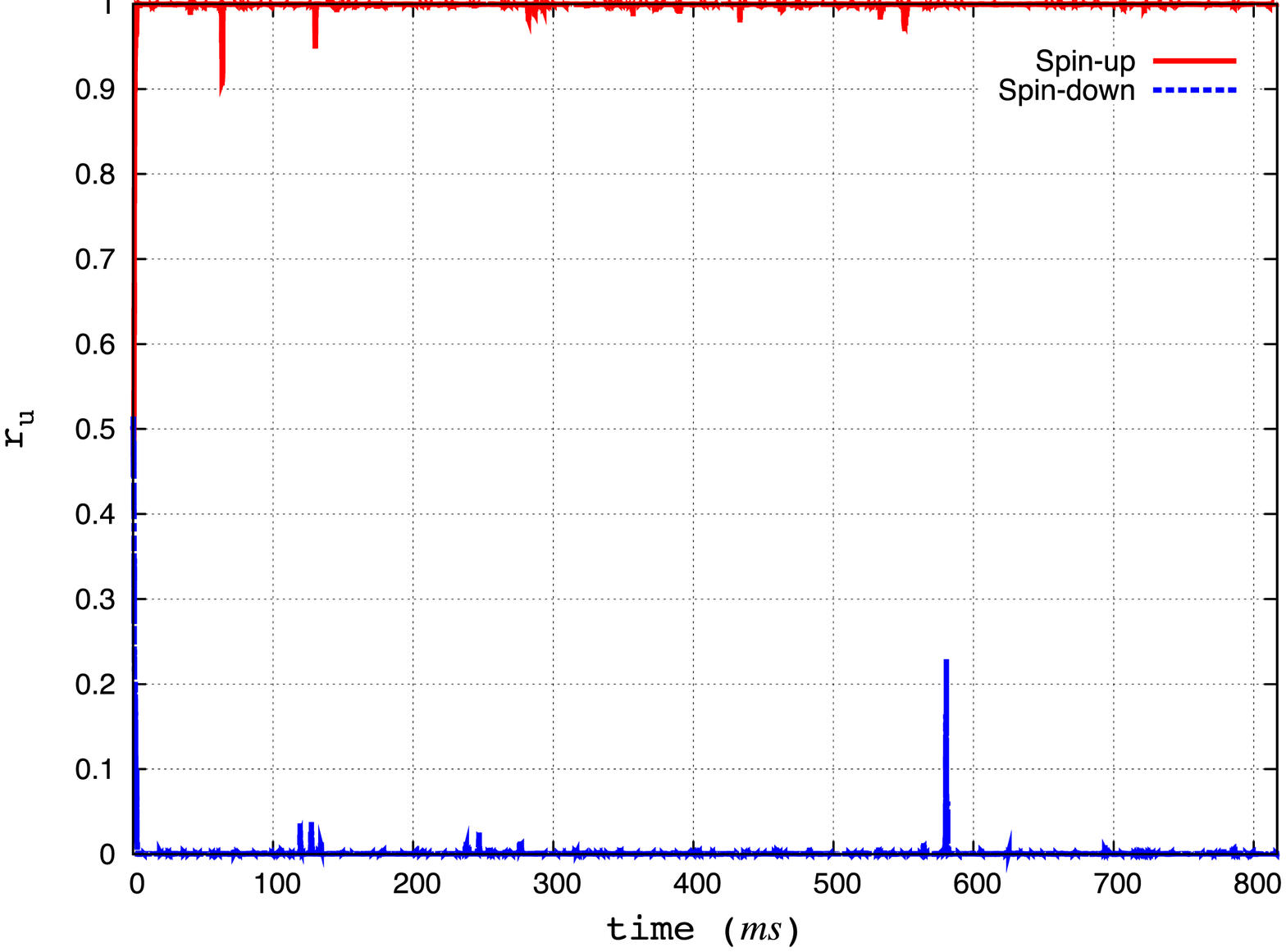}} &
    	 \subfigure[]{\label{fig:freqShift} \includegraphics[width = 3.0in, height=2.5in]{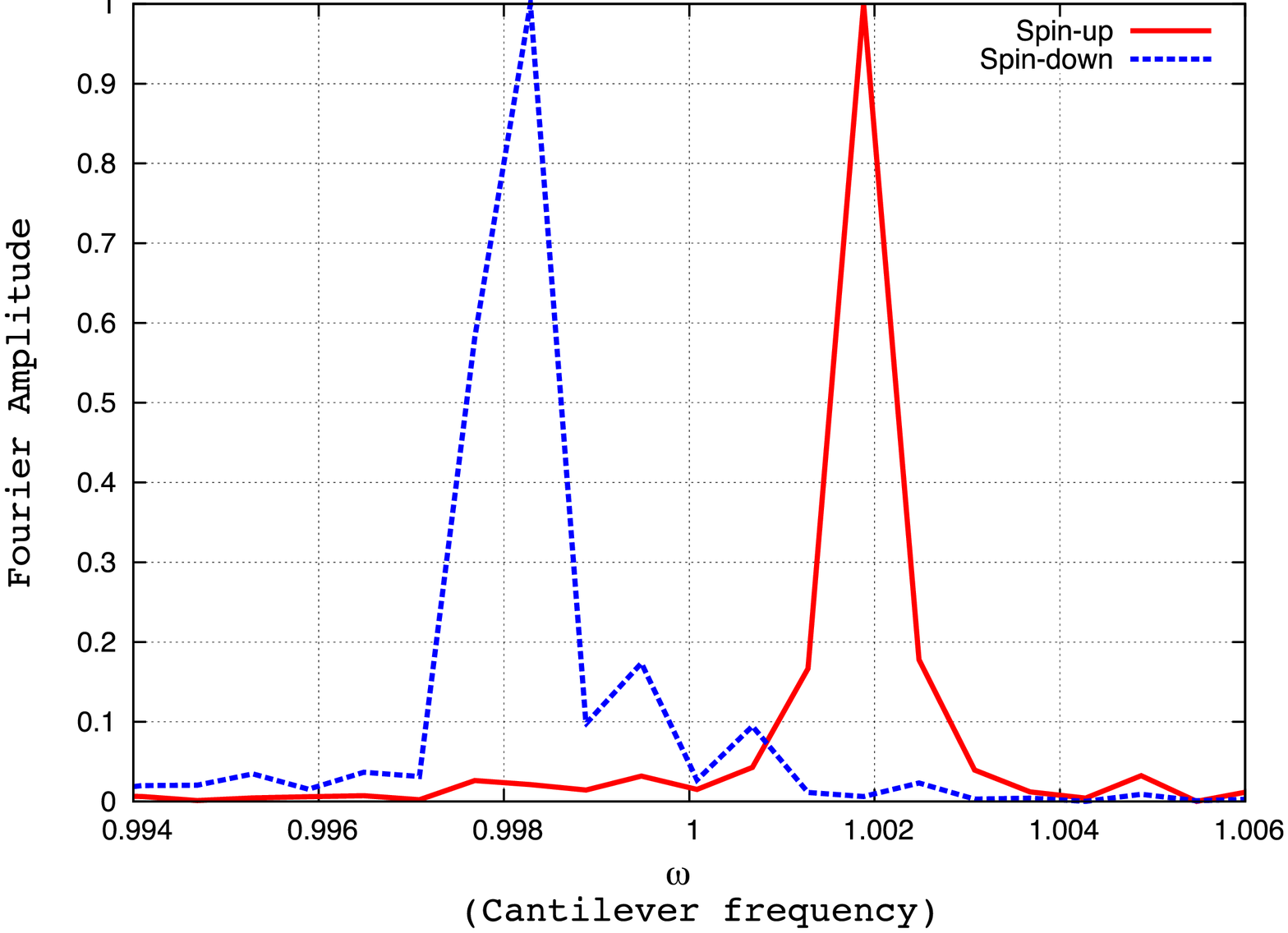}} \\
      \end{tabular}
  \end{center}
  \caption{(Color online.) Frequency shift in OSCAR MRFM. We show the frequency shift in the OSCAR MRFM protocol, for two different trajectories. Figure~\ref{fig:spinprob_2trajs} shows the time evolution of the spin-up probability, $r_u$, for the two trajectories; in the first, the spin relaxes to its up state, while in the second, it relaxes to its down state. Figure~\ref{fig:freqShift} plots the Fourier amplitude (in arbitrary units) as a function of cantilever frequency, $\omega$, corresponding to the two trajectories in Fig.~\ref{fig:spinprob_2trajs}. The spin noise rate $\kappa_s$ used in this plot is $10^{-5}$. The Fourier amplitude is calculated using a standard FFT algorithm; the number of samples is $N=2^{19}$, and the sample spacing is $\Delta t=0.02$.}
  \label{fig:oscar_freqShift}
\end{figure*}

The value of sampling period, $T_{Sampling}$, in Eq.~(\ref{eqn:sample_period}) is in dimensionless units. In our calculation of parameter values for our simulations, we assumed that the cantilever frequency, in physical units, is
\[
f_c^{phy} = \omega_{c}^{phy}/2 \pi \approx 16 \, \text{kHz}.
\]
Using this frequency, the duration of sampling, in physical units, is 
\[
T_{Sampling}^{phy} \: = \: T_{Sampling}/f_{c}^{phy} \: \approx \: 656 \, \text{ms}, 
\]
and the shift in the cantilever frequency is 
\[
\Gamma^{phy} \: = \: \Gamma \times f_c^{phy} \: \approx \: 29 \, \text{Hz}.
\]
Thus, it takes about $650 \: ms$ for the OSCAR protocol to determine a shift of around $30 \: Hz$ in cantilever frequency, and consequently, to ascertain the orientation of the spin. The time scale of the spin noise must be longer than the sampling duration to use OSCAR MRFM as a single-spin measurement device.

%\\\\\\\\\\\\\\\\\\\\\\\\\\\\\\\\\\\\\\\\\\\\\\\\\\\\\\\\\\\\\\\\\\\\\\\\\\\\\\\\\\\\\\\\\\\\\\\\\\\\\\\\\\\\\\\\\\\\\\\\\\\\\\\\\\\\\\\\\\\\\\\\\\\\\\\\\\\\\\\\\\\\\\\\\\\\\\\\\\\\\\\\
%
\section{Conclusions}
\label{sec:conclusions}

In this work, we modeled the OSCAR MRFM operation including decoherence effects on the cantilever (both thermal and monitoring), as well as spin noise due to magnetic sources. We describe the evolution of the system using a quantum stochastic master equation, the most general description in a Markovian framework. We then simplified this description using a series of approximations, and arrived at a semi-classical description of the system based on a Gaussian approximation of the cantilever state. 

We numerically compared the Gaussian approximation to the fully quantum stochastic master equation, and found that the Gaussian approximation tracked the fully quantum stochastic master equation very closely. Thus, we conclude that the Gaussian approximation of the cantilever is useful for our range of parameter values, and that the MRFM system implementing the OSCAR protocol can be described by a closed set of $11$ coupled stochastic differential equations.

Finally, we used the Gaussian approximation equations to numerically verify OSCAR as a useful single-spin measurement protocol. The critical element in the OSCAR protocol is the time (sampling) it takes to determine the shift in resonant cantilever frequency. For the parameter values chosen in our simulations, we found that the sampling duration is around $650$ ms, and the frequency shift is about $30$ Hz. The sampling duration thus places a bound on the time scale of spin noise that  the OSCAR protocol, implementing a single-spin measurement, can tolerate. 

%\\\\\\\\\\\\\\\\\\\\\\\\\\\\\\\\\\\\\\\\\\\\\\\\\\\\\\\\\\\\\\\\\\\\\\\\\\\\\\\\\\\\\\\\\\\\\\\\\\\\\\\\\\\\\\\\\\\\\\\\\\\\\\\\\\\\\\\\\\\\\\\\\\\\\\\\\\\\\\\\\\\\\\\\\\\\\\\\\\\\\\\\
\section*{Acknowledgments}
TAB and SR were supported in part by NSF Grant No.~ECS-0507270 and NSF CAREER Grant No.~CCF-0448658.
H.S.G. acknowledges support from the National Science
Council, Taiwan, under Grant No. 97-2112-M-002-012-MY3, 
support from the Frontier and Innovative Research Program 
of the National Taiwan University under Grants Nos. 97R0066-65 and 
97R0066-67, and support from the focus group
program of the National Center for Theoretical Sciences, Taiwan.
%\\\\\\\\\\\\\\\\\\\\\\\\\\\\\\\\\\\\\\\\\\\\\\\\\\\\\\\\\\\\\\\\\\\\\\\\\\\\\\\\\\\\\\\\\\\\\\\\\\\\\\\\\\\\\\\\\\\\\\\\\\\\\\\\\\\\\\\\\\\\\\\\\\\\\\\\\\\\\\\\\\\\\\\\\\\\\\\\\\\\\\\\
% Bibliography
%


\begin{thebibliography}{10}
\bibitem{SME-Package}{This package was developed by Todd Brun and Bilal Shaw, and is an extension of QSD library developed by Shack and Brun~\cite{Shack97}.}
\end{thebibliography}
\end{document}